\documentclass[12pt,oneside]{article}
\newcommand{\be}{\begin{equation}}
\newcommand{\ee}{\end{equation}}

\def\bea{\begin{eqnarray}}
\def\eea{\end{eqnarray}}
\def\bean{\begin{eqnarray*}}
\def\eean{\end{eqnarray*}}
\def\thru#1{\mathrel{\mathop{#1\!\!\!\!/}}}
\def\pa#1{\frac{\partial}{\partial z^{}_#1}}
\def\pab#1{\frac{\partial}{\partial \overline z^{}_#1}}
\def\part#1{\partial^{}_#1}
\def\partb#1{\overline\partial^{}_#1}

\begin{document}
\thispagestyle{empty}
\setcounter{page}{0}
\renewcommand{\theequation}{\thesection.\arabic{equation}}

{\hfill{\tt hep-th/0203008}}

\vspace{2cm}

\begin{center}
{\bf EULER MULTIPLETS AND LIGHT-CONE SUPERSYMMETRY}

\vspace{1.4cm}

LARS BRINK

\vspace{.2cm}

{\em Department of Theoretical Physics} \\
{\em Chalmers University of Technology} \\
{\em S-412 96 G\"oteborg, Sweden}
\end{center}

\vspace{-.1cm}

\centerline{{\tt Lars.Brink@fy.chalmers.se}}

\vspace{1cm}

\centerline{ABSTRACT}

\vspace{- 4 mm}  

\begin{quote}\small
In a preceding report Pierre Ramond outlined a programme we have followed recently. An important ingredient in a future unique theory for basic physics must be a unique mathematical structure. It is then interesting to investigate the underlying mathematical structure of 11-dimensional space-time. A search for exceptional group structures specific to eleven dimensions leads us then to a study of the  coset $F_4/SO(9)$,which  may provide  a generalization of eleven-dimensional supergravity. This scheme is more general though and in this report we will show the corresponding structure related to the coset $SU(3)/SU(2) \times U(1)$, where we interpret the $U(1)$ as the helicity group in four dimensions.
\end{quote}

\baselineskip18pt

\newpage

\setcounter{equation}{0}
\section{Introduction}
String theory has given us an understanding of the dimension of spacetime as a crucial concept for building consistent 
quantum mechanical theories. This insight has been  further strengthened with the introduction of supersymmetry, since
the size of spinors depends not only on the size of spacetime but also
on the possibility to implement the Majorana and the Weyl conditions.
The little group for the superstring is $SO(8)$. This is really one of the
most beautiful and unique groups. Its Dynkin diagram is the famous Mercedes symbol,
and the group is really the Mercedes of the orthogonal groups. It has a
triality symmetry and its three eight-dimensional  vector, 
spinor and  cospinor representations are readily interchangeable. In the light cone
formulation of the superstring it is these three representations that
build up the theory and in quantum calculations there are marvellous
cancellations between the bosonic and fermionic contributions that
render the theory perturbatively finite.

From this viewpoint the emergence of M-theory as an even more general
theory was unexpected. Can one find some group-theoretic reasons why an
eleven-dimensional spacetime should make special sense. The classic
reason  is that it is the maximal one to carry a supermultiplet
with the graviton as the highest spin field. Can one argue for it from the point of view
of the little group $SO(9)$? Its Dynkin diagram does not carry the same
beauty as that of $SO(8)$. There is no symmetry in it and we know of no cars using it as a symbol, not even the
Trabant. (We apologize to our good friends in Eastern Germany). However, $SO(9)$ is also a beautiful group in the sense of having
a wealth of internal relations between the irreps, again making it a
strong candidate for a fundamental symmetry. It was discovered by Ramond~\cite{P1} that some of its irreps naturally group together into triplets which are such
that bosonic and fermionic degrees of freedom match up. In
fact there is an infinity of triplets with identical properties to the smallest one, which is the supergravity multiplet of
eleven dimensions. This makes the triplets interesting not only from a
mathematical point of view but also from a quantum physical one. In a
higher order loop calculation in a supersymmetric theory there are huge
cancellations between the fermionic and the bosonic contributions since
their contributions can be written in terms of group theoretic indices
related to the little group which match up~\cite{Curtwrong}. Exactly the same kind of
matching occurs in fact for all the triplets, making them candidates for  
a quantum theory extending supergravity.

The higher triplets involve higher spin massless fields. A problem in string
theory not fully understood is what happens in the limit of infinite Regge slope, i.e. the zero-tension limit. Here all fields will
be massless and will have arbitrarily high spins. No natural solution to this problem has been found in ten dimensions so it
is interesting to ask if one could find a natural theory in eleven
dimensions with massless fields of arbitrary spin. We now have
candidates among the triplets mentioned above. An interesting question
is what happens to the symmetries of a massive theory when all particles become massless.
Usually the linearly realized symmetry is enhanced as it happens when a
spontaneously broken symmetry is restored. The gauge invariance will be restored as usual but what would happen to the
(spacetime) supersymmetry of the string? It would not be changed if it is a global symmetry, but for higher spins it will be a local symmetry coupled to reparametrization and to higher spin field symmetry so we expect it to be part of a  restored symmetry. Will it be enlarged? We do not know but it seems likely that, if there is a fundamental theory for the massless fields, they should all be connected. If we treat this theory in the light cone gauge all symmetries should be global and the only supersymmetry that could connect an infinite representation irreducibly occurs when the order N of the supersymmetry goes to infinity. In the sequel we will show another kind of symmetry and infinite irreps that are connected to the triplets above  which is a generalization of supersymmetry.

Since the discovery of the triplets there has been great progress in the
mathematical understanding of their beautiful properties \cite{GKRS}. The punch line
is that this phenomenon occurs when there is a subgroup with the same
rank as the full group. In the case above $SO(9)$, alias $B_4$, is a subgroup
of $F_4$ with the same rank, and the quotient space $F_4/B_4$ has Euler
number three giving a triplet of $SO(9)$ to every irrep of  $F_4$.   We have elsewhere \cite{BR,BR2,PR2}
listed the cases with up to 16-dimensional cosets and in this list we
find multiplets with the above properties, extending the multiplets of
besides the 11-dimensional supergravity also  $N=8$ supergravity, $N=4$ Yang-Mills
and the $N=2$ hypermultiplet. There is a beautiful equation due to Kostant~\cite{KOS}
which is a Dirac-like equation on the coset, which has as solutions the
multiplets.

It is quite interesting that the exceptional algebra $F_4$ enters into
the description of an 11-dimensional theory. We have seen the
exceptional groups emerge as gauge groups and there is a direct line
from the gauge group of the Standard Model via $SU(5) = E_4$ and $SO(10)
= E_5$ up to the ultimate exceptional group $E_8$, the Volvo of the groups. So far there has not
been any trace of exceptional groups extending the space time symmetry.
There is a simple reason for this since they relate tensor and spinor
representations of their orthogonal subgroups, while spin statistics
treat them differently. However, the exceptional groups are the most
unique and beautiful ones and it is many physicists' dream that they represent
the ultimate symmetry of the world. In the preceding report Pierre Ramond outlined our study of the case $F_4/SO(9)$. In this report I will then discuss in quite some detail the simpler case of $SU(3)/SU(2) \times U(1)$, which we interpret as a model for higher spin massless fields in four dimensions, generalizing the $N=2$ hypermultiplet.

One question that will arise is if we should use all triplets as a candidate
theory or if there is a natural selection among them. In this report we show a method where we define a superfield which naturally has an
equal number of bosonic and fermionic degrees of freedom and is a
natural extension of the superfield containing the smallest multiplet. We will use a light-cone  frame formulation which
will tie together the superspace with the external symmetry. We will
show that the infinite superfield which is a solution to Kostant's
equation is a representation of a new infinitely extended superalgebra.

\setcounter{equation}{0}
\section{Euler Triplets for $SU(3)/SU(2)\times U(1)$}
In the following, we present a detailed analysis of the Euler triplets 
associated with the coset $SU(3)/SU(2)\times U(1)$. There is an infinity of 
Euler triplets which are solutions of Kostant's equation associated with the coset. 
The most trivial solution describes the light-cone degrees of freedom of 
the $N=2$ hypermultiplet
in four dimensions, when the $U(1)$ is interpreted as helicity. Hence 
we begin by reminding the reader of the 
well-known light-cone description of that multiplet.  

\subsection{The $N=2$ Hypermultiplet in 4 Dimensions}
The massless $N=2$ scalar hypermultiplet contains two Weyl spinors 
and two complex scalar fields, on which   the $N=2$ SuperPoincar\'e 
algebra is realized. Introduce the  light-cone Hamiltonian

\be
P_{}^-=\frac{p\overline p}{p^+}\ ,\ee
where 
$
p=\frac{1}{\sqrt{2}}(p_{}^1+ip_{}^2)\nonumber \ .$ The front-form 
supersymmetry generators satisfy the anticommutation relations

\bea \nonumber \{{\cal Q}^{m}_+ ,\overline {\cal Q}^{n}_+\}&=&-
2\delta_{}^{mn}p_{}^+\ ,\\
\{{\cal Q}^{m}_- ,\overline {\cal Q}^{n}_-\}&=&-2\delta_{}^{mn}
\frac{p\overline p}{p_{}^+}\ ,~~~~~m,n=1,2\ ,\\
\nonumber \{{\cal Q}^{m}_+ ,\overline {\cal Q}^{n}_-\}&=&-2p\delta_{}^{mn}\ .\eea
The kinematic supersymmetries are expressed as 

\be
{\cal Q}^{m}_+=-\frac{\partial}{\partial\overline\theta^m}-\theta_m p_{}^+\ ,\qquad
\overline {\cal Q}^{m}_+=\frac{\partial}{\partial\theta^m}+\overline\theta_m p_{}^+\ ,\ee
while the kinematic Lorentz generators are given by

\bea
\label{M}
M_{}^{12}&=&i(x\overline p-\overline xp)+\frac{1}{2}\theta_m
\frac{\partial}{\partial\theta_m}-\frac{1}{2}\overline\theta^m
\frac{\partial}{\partial\overline\theta^m}\ ,\\ \nonumber   
M_{}^{+-}&=&-x_{}^-p_{}^+ -\frac{i}{2}\theta_m\frac{\partial}
{\partial\theta_m}-\frac{i}{2}\overline\theta^m\frac{\partial}
{\partial\overline\theta^m}\ ,   
\\ \nonumber
M_{}^{+}&\equiv&\frac{1}{\sqrt{2}}(M_{}^{+1}+iM_{}^{+2})=
-xp_{}^+\ ,\qquad 
\overline M_{}^{+}=-\overline xp_{}^+\ ,\\ \nonumber
\eea
where $x=\frac{1}{\sqrt{2}}(x_{}^1+ix_{}^2)$, and where the two complex Grassmann variables satisfy
the anticommutation relations
\bea
\nonumber \{\theta_m,\frac{\partial}{\partial\theta_n} \}&=&
\{\overline\theta^m,\frac{\partial}{\partial\overline\theta^n}\}=\delta^{mn}\ ,\\ 
\nonumber \{\theta_m,\frac{\partial}{\partial\overline\theta^n} \}&=&
\{\overline\theta^m,\frac{\partial}{\partial\theta_n} \}
=0\ .\eea
The  (free) Hamiltonian-like supersymmetry generators are simply
\be
{\cal Q}^{m}_-=\frac{\overline p}{p_{}^+}{\cal Q}^{m}_+\ ,\qquad
\overline {\cal Q}^{m}_-=\frac{ p}{p_{}^+}\overline {\cal Q}^{m}_+\ ,\ee
and the light-cone boosts are given by
\bea
M_{}^{-}&=&x_{}^-p-\frac{1}{2}\{x,P_{}^-\}+i\frac{p}{p_{}^+}\theta_m
\frac{\partial}{\partial\theta_m}\ ,
\\ \nonumber
\overline M_{}^{-}&=&x_{}^-\overline p-\frac{1}{2}\{\overline x,P_{}^-\}+i\frac{
\overline p}{p_{}^+}\overline \theta^m
\frac{\partial}{\partial\overline \theta^m}\ .\eea

This representation of the superPoincar\'e algebra is reducible, as it 
can be seen to act on reducible superfields $\Phi(x^-,x^i,\theta_m,\overline\theta^m)$, because  the operators 

\be
{\cal D}_+^m~=\frac{\partial}{\partial\overline\theta^m}-\theta_m p_{}^+\ ,\ee
 anticommute with the supersymmetry generators. As a result, one can 
achieve irreducibility by acting on superfields for which 
\be
\label{Ch}
{\cal D}_+^m~\Phi=[\frac{\partial}{\partial\overline\theta^m}-\theta_m p_{}^+]\Phi =0\ ,\ee
solved by the chiral superfield

\be
\label{Phi}
\Phi(y^-,x^i,\theta_m)= \psi^{}_0(y^-,x^i) + \theta^{}_m \psi_{}^m(y^-,x^i) + 
\theta^{}_1 \theta^{}_2\psi_{}^{12}(y^-,x^i)\ .\ee
The field entries of the scalar hypermultiplet now depend on the combination

\be
y^-=x^--i\theta_m\overline\theta^m\ ,\ee
and the transverse variables.  Acting on this chiral superfield, the 
constraint is equivalent to requiring that

\be
{\cal Q}_+^m\approx -2p^+\theta_m\ ,\qquad \overline{\cal
Q}_+^m\approx 
\frac{\partial}{\partial\theta_m}\ ,\ee
where the derivative is meant to act only on the naked $\theta_m$'s, 
not on those hiding in  $y^-$. This light-cone representation is 
well-known, but we repeat it here to set our conventions and notations.

\subsection{Coset Construction}
Let $T^A$ , $A=1,2,\dots 8$, be the generators of $SU(3)$. Among
those, $T^i$, $i=1,2,3$,  and $T^8$   generate its $SU(2)\times U(1)$
subalgebra. Introduce  Dirac matrices over the coset 

$$ \{\gamma^a, \gamma^b \} = 2\delta ^{ab}\ ,$$
for $a,b=4,5,6,7$, to define 
 the Kostant equation over the coset $SU(3)/SU(2)\times U(1)$ as

\be
\thru {\cal K}\Psi~=~\sum_{a=4,5,6,7}\gamma_{}^aT^{}_a\,\Psi~=~0\ .\nonumber \ee
 The Kostant operator commutes with  the $SU(2)\times U(1)$ generators

\be
L^{}_i=T^{}_i+S^{}_i\ ,~i=1,2,3\ ;\qquad L^{}_8=T^{}_8+S^{}_8\ .\ee
These are a sum of the $SU(3)$ generators and the $``$spin" part, expressed in terms 
of the $\gamma$ matrices as 

\be 
S^{}_j=- \frac{i}{4}f^{}_{jab}\gamma_{}^{ab}\ , \qquad S^{}_8=-
\frac{i}{4}f^{}_{8ab}\gamma_{}^{ab}\ ,\ee
where $\gamma^{ab}=\gamma^a\gamma^b\ ,a\ne b\ ,$ and $f^{}_{jab}\ ,f^{}_{8ab}$ are structure functions of $SU(3)$.

The Kostant equation has an infinite number of solutions which come in groups of three representations of 
$SU(2)\times U(1)$, called Euler triplets. For each representation of
$SU(3)$, there is a unique Euler triplet, each given by three representations

$$\{a_1,a_2\}~\equiv~[a^{}_2]_{-\frac{2a^{}_1+a_2+3}{6}}\oplus 
[a^{}_1+a_2+1]_{\frac{a^{}_1-a_2}{6}}\oplus [a^{}_1]_{\frac{2a^{}_2+a_1+3}{6}}\ ,$$
where $a_1,a_2$ are the Dynkin labels of the associated $SU(3)$ representation. 
Here, $[a]$ stands for the $a=2j$ representation of $SU(2)$, and the
subscript denotes the $U(1)$ charge. The Euler triplet corresponding to $a_1=a_2=0$,

$$\{0,0\}~=~[0]_{-\frac{1}{2}}\oplus [1]_{0}\oplus [0]_{\frac{1}{2}}\ ,$$
describes the degrees of freedom of the $N=2$ supermultiplet,
where the properly normalized $U(1)$ is interpreted as the helicity of the four-dimensional Poincar\' e algebra. 

Below, we wish to explore the  possibility of linking this 
supersymmetric triplet to those for which $a_{1,2}\ne 0$, 
while preserving at least relativistic invariance. Of particular interest will be the algebraic operations that link the different Euler triplets. Their use will enable us to define supersymmetry-like  operations acting on the higher Euler triplets, which serve as the shadow of the light-cone supersymmetry of the lowest Euler triplet.

The $U(1)$ 
charges of the higher triplets are  generally rational numbers, which means that they display parastatistics, but the triplets for which  

$$a_1~=~a_2\ ,~~~~~~{\rm mod}~(3)\ ,$$
contain half-odd integer or integer $U(1)$ charges, and satisfy Fermi-Dirac statistics. A self-conjugate subset  

$$\{a,a\}~:~~~~[a]_{-\frac{a+1}{2}}\oplus [2a+1]_{0}\oplus [a]_{\frac{a+1}{2}}\ ,$$
 contains equal number of half-odd integer-helicity fermions and
integer-helicity bosons, and satisfy CPT.  As the helicity gap between the representations increases indefinitely in half integer steps, the symmetry operations that relate its members have helicities $\pm(a+1)/2$. When $a=0$,  they can be identified with the usual supersymmetries, and they are fermionic as long as $a$ is even.

The others are generated from  complex representations with  $N=1,2,\dots$, of the form

$$\{a,a+3N\}~:~~~~[a]_{-\frac{a+N+1}{2}}\oplus
[2a+3N+1]_{\frac{N}{2}}
\oplus [a+3N]_{\frac{a+N+1}{2}}\ .$$ 
The helicity gap also increases in half integer steps, starting at one-half. 
Since  CPT requires states of opposite helicity, these must be 
accompanied by their conjugates, 
 $\{a+3N,a\}$, with all helicities reversed. 

A special case deserves consideration: when $a=0$, the helicity gap can be as small as $1/2$, like the regular supersymmetry. The simplest example is

$$\{0,3\}~:~~~~[0]_{-1}\oplus
[4]_{\frac{1}{2}}
\oplus [3]_{1}\ ,$$
where the helicity gap is $3/2$ and $1/2$. When we add the $CPT$ conjugate
$$ \{3,0\}~:~~~~   \oplus [0]_{1}\oplus
[4]_{-\frac{1}{2}}
\oplus [3]_{-1}\ ,$$
  we end up with states separated by half a unit of helicity as in the supersymmetric multiplets. As they occur in different representations of $SU(2)$,  equality between bosons and fermions is achieved only after including the CPT conjugate. Unlesss the $SU(2)$ is broken, relativistic supersymmetry cannot be implemented on these states. The case $N=2$ yields states of helicity $1$ and $3/2$, and $N=3$ contains eleven states of helicity $2$, and so on.

It appears that while Poincar\'e symmetry can be implemented on an infinite subset of Euler triplets, relativistic supersymmetry can be realized on a finite subset, and only after the $SU(2)$ is broken. In particular the need for operators that shift helicity by more than half units makes it unlikely that a relativistic supersymmetric theory of Euler triplets can be found. 

In addition, the higher Euler triplets include states 
with  helicities larger than $2$, which cannot be interpreted 
as massive relativistic states since they do not arrange themselves in 
$SO(3)$ representations. Hence they  must
be viewed as massless particles in four dimensions, leading 
to a theory of massless states of spin higher than $2$. 

There are well-known difficulties with such theories. In particular, 
they do not have covariant  energy momentum tensors, and  in the flat space limit they must decouple from the gravitational 
sector, although these no-go theorems do not apply if  there 
is an infinite number of particles, and there exist theories which circumvent them. 
Our purpose is to investigate if a relativistic theory can be formulated 
with an infinite number of Euler multiplets, in which a light-cone version of a new type of space-time fermionic symmetry is present. 

\subsection{Grassmann Numbers and Dirac Matrices}
In order to use the superfield technique we will identify the spin part of the $U(1)$ generator $S^{}_8$ with the spin part in Equ.~(\ref{M}) taking the condition (\ref{Ch}) into account. This will mean that we write also $S^{}_i$ in terms of the $\theta's$. An appropriate representation is then

\bea
\gamma_{}^4+i\gamma_{}^5&=&i\sqrt{\frac{2}{p^+}}{\cal Q}_+^1\ ,\qquad
\gamma_{}^4-i\gamma_{}^5=i\sqrt{\frac{2}{p^+}}\overline{\cal Q}_+^1\\
\gamma_{}^6+i\gamma_{}^7&=&i\sqrt{\frac{2}{p^+}}{\cal Q}_+^2\ ,\qquad
\gamma_{}^6-i\gamma_{}^7=i\sqrt{\frac{2}{p^+}}\overline{\cal Q}_+^2\ ,
\eea
in terms of the kinematic $N=2$ light-cone supersymmetry generators 
defined in the previous section. We can check that $S_8$ indeed agrees with the spin part of Equ.~(\ref{M}) (after proper normalization). As the Kostant operator 
anticommutes with the constraint operators

\be\{~\thru {\cal K},~{\cal D}_+^m\}~=~0\ ,\ee
 its solutions can be written as chiral superfields, on which the $\gamma$'s become   

 
\bea
\gamma_{}^4+i\gamma_{}^5&=&-2i\sqrt{2p^+}~\theta_1\ ,\qquad
\gamma_{}^4-i\gamma_{}^5=i\sqrt{\frac{2}{p^+}}~\frac{\partial}{\partial\theta_1}\\
\gamma_{}^6+i\gamma_{}^7&=&-2i\sqrt{2p^+}~\theta_2\ ,\qquad
\gamma_{}^6-i\gamma_{}^7=i\sqrt{\frac{2}{p^+}}~\frac{\partial}{\partial\theta_2}\ ,
\eea
The complete $``$spin" parts of the $SU(2) \times U(1)$ generators,  expressed
in terms of  Grassmann variables, do not depend on $p^+$,

\bea
S^{}_1&=&\frac{1}{2}  ( \theta^{}_1\frac{\partial}{\partial
\theta^{}_2}+ 
\theta^{}_2\frac{\partial}{\partial \theta^{}_1})\ ,\qquad
S^{}_2=-\frac{i}{2}  (\theta^{}_1 \frac{\partial}{\partial
\theta^{}_2}-  
\theta^{}_2 \frac{\partial}{\partial \theta^{}_1})\cr
S^{}_3&=&\frac{1}{2}  (\theta^{}_1\frac{\partial}{\partial
\theta^{}_1}-  
\theta^{}_2\frac{\partial}{\partial \theta^{}_2})\ ,\qquad
S^{}_8=\frac{\sqrt3}{2} (\theta^{}_1 \frac{\partial}{\partial
\theta^{}_1}+
\theta^{}_2 \frac{\partial}{\partial \theta^{}_2}-1)\ .
\eea
Using  Grassmann properties, the $SU(2)$ Casimir operator can be written as  

\be\vec{S}^2= \frac{3}{4}  ( \theta^{}_1\frac{\partial}{\partial
\theta^{}_1}  
- \theta^{}_2 \frac{\partial}{\partial \theta^{}_2})^2\ ;\ee
it has only two eigenvalues, $3/4$ and zero. These $SU(2)$ generators obey a simple algebra

\be
S^{}_i\,S^{}_j=\frac{1}{3}\vec S\cdot \vec S\,\delta^{}_{ij}+\frac{i}{2}\epsilon^{}_{ijk}S^{}_k\ .\ee
The helicity, identified with $S_8$ up to a normalizing factor
of $\sqrt{3}$, leads to half-integer helicity values on the
Grassmann-odd components of the (constant) superfield representing the
hypermultiplet. 

\subsection{Linear Realization of $SU(3)$}
The $SU(3)$ generators can be conveniently expressed on three complex
variables and their conjugates. Define for convenience the differential operators

$$\part1\equiv\pa1\ ,\qquad \partb1\equiv\pab1\ ,~{\rm etc.}\ ,$$
in terms of which the generators are given by 

$$ T^{}_1+iT^{}_2=z^{}_1\part2-\overline z^{}_2\partb1\ ,\qquad 
T^{}_1-iT^{}_2=z^{}_2\part1-\overline z^{}_1\partb2\ ,$$
$$T^{}_4+iT^{}_5=z^{}_1\part3-\overline z^{}_3\partb1\ ,\qquad 
T^{}_4-iT^{}_5=z^{}_3\part1-\overline z^{}_1\partb3\ ,$$
$$ T^{}_6+iT^{}_7=z^{}_2\part3-\overline z^{}_3\partb2\ ,\qquad 
T^{}_6-iT^{}_7=z^{}_3\part2-\overline z^{}_2\partb3\ ,$$
and

$$ T^{}_3=\frac{1}{2}(
z^{}_1\part1-z^{}_2\part2-\overline z^{}_1\partb1+\overline z^{}_2\partb2)\ ,$$
$$ T^{}_8=\frac{1}{2\sqrt{3}}(
z^{}_1\part1+z^{}_2\part2-\overline z^{}_1\partb1-\overline z^{}_2\partb2-2
z^{}_3\part3+2\overline z^{}_3\partb3)\ .$$
These  act as hermitian operators on holomorphic functions of
$z_{1,2,3}^{}$ 
and $\overline z_{1,2,3}^{}$, normalized with respect to the inner product

$$ (f,g)\equiv~\int d^3zd^3\overline z~e^{-\sum_i\vert z_i\vert^2}~
f^*(z,\overline z)~g(z,\overline z)\ .$$
Acting on the highest-weight states, the $SU(3)$ quadratic Casimir operator is 

$$ C_2^{SU(3)}\equiv\sum_{a=1}^8T^{}_aT^{}_a\Big\vert_{\rm
highest~weight}=
T^{}_3(T^{}_3+1)+T^{}_8(T^{}_8+\sqrt{3})\ .$$
The second Casimir operator is cubic, and of no concern here. Rather
than labelling the representations in terms of their eigenvalues, it is
more convenient to introduce the 
positive integer Dynkin labels $a_1$ and $a_2$. We have

$$ T^{}_3\vert~ a_1,a_2>=~\frac{a_1}{2}\vert~ a_1,a_2>\ ,\qquad
T^{}_8\vert~ 
a_1,a_2>~=~\frac{1}{2\sqrt{3}}(a_1+2a_2)\vert~ a_1,a_2>\ ,$$
so that

$$ C_2^{SU(3)}~=~(a^{}_1+a^{}_2)+\frac{1}{3}(a_1^2+a_1^{}a_2^{}+a^2_2)\ .$$
The highest-weight states of each $SU(3)$ representation are holomorphic 
polynomials of the form

$$ z^{a_1}_1\overline z_3^{a_2}\ ,$$
where $a_1,a_2$ are its  Dynkin indices:  all
representations 
of $SU(3)$ are homogeneous holomorphic polynomials. We further note 
that any function of  the quadratic invariant

$$ Z^2\equiv \vert z_1^{}\vert^2+\vert z_2^{}\vert^2
+\vert z_3^{}\vert^2\ ,$$
 can  multiply these polynomials without affecting their $SU(3)$ 
transformation properties.

 Finally we note that the Casimir operator of the $SU(2)$ subalgebra is given by

\be\vec T\cdot\vec T~=~\frac{1}{4}D_\perp(D_\perp+2)\ ,\ee
where

$$D^{}_\perp~=~z^{}_1\part1+z^{}_2\part2+\overline z^{}_1\partb1+\overline z^{}_2\partb2\ ,$$
so that the spin of the $SU(2)$ representation is simply

\be
J~=~\frac{1}{2}D^{}_\perp\ .\ee

\subsection{Solutions of Kostant's Equation}
Consider now  Kostant's equation

$$\thru {\cal K}\Psi~=~\sum_{a=4,5,6,7}\gamma_{}^aT^{}_a\Psi~=~0\ .$$
Expanding the solutions and the Dirac matrices in terms of Grassmann 
variables yields two independent pairs of equations

$$(T^{}_4+iT^{}_5)\psi^{}_1+(T^{}_6+iT^{}_7)\psi^{}_2~=~0\ ;\qquad 
(T^{}_4-iT^{}_5)\psi^{}_2-(T^{}_6-iT^{}_7)\psi^{}_1~=~0\ ,$$
 and

$$(T^{}_4-iT^{}_5)\psi^{}_0-(T^{}_6+iT^{}_7)\psi^{}_{12}~=~0\ ;\qquad 
(T^{}_6-iT^{}_7)\psi^{}_0+(T^{}_4+iT^{}_5)\psi^{}_{12}~=~0\ ,$$
that is

$$(z^{}_1\part3-\overline z^{}_3\partb1)\psi^{}_1+(z^{}_2\part3-
\overline z^{}_3\partb2)\psi^{}_2~=~0\ ;\qquad (z^{}_3\part1-
\overline z^{}_1\partb3)\psi^{}_2-(z^{}_3\part2-\overline z^{}_2
\partb3)\psi^{}_1~=~0\ ,$$
$$(z^{}_3\part1-\overline z^{}_1\partb3)\psi^{}_0-(z^{}_2\part3-
\overline z^{}_3\partb2)\psi^{}_{12}~=~0\ ;\qquad (z^{}_3\part2-
\overline z^{}_2\partb3)\psi^{}_0+(z^{}_1\part3-\overline z^{}_3\partb1)\psi^{}_{12}~=~0\ .$$
The homogeneity operators

$$D=z^{}_1\part1+z^{}_2\part2+z^{}_3\part3\ ,\qquad \overline 
D=\overline z^{}_1\partb1+\overline z^{}_2\partb2+ \overline z^{}_3\partb3\,$$
commute with $\thru {\cal K}$, allowing the solutions of  Kostant
equation to  be arranged 
in terms of homogeneous polynomials, on which  $a_1$ is the eigenvalue of $D$ and $a_2$ that of $\overline D$.
The solutions can also be labeled in terms of the $SU(2)\times U(1)$ generated by the operators

$$L^{}_i=T^{}_i+S^{}_i\ ,~i=1,2,3\ ;\qquad L^{}_8=T^{}_8+S^{}_8\ .$$
The solutions for each triplet, conveniently written explicitly only for the highest weight states, are of the form 

\bea 
\Psi&=& z_3^{a_1}~\overline z_2^{a_2}\ ,~~~~~~~~~~{\rm labels}~~[a_2]_{-\frac{2a^{}_1+a_2+3}{6}}
\ ,\cr 
&+& \theta_1~z_1^{a_1}~\overline z_2^{a_2}\ ,~~~~~~{\rm labels}~~
[a_1+a_2+1]_{\frac{a^{}_1-a_2}{6}}\ ,\cr
&+& \theta_1\theta_2~z_1^{a_1}~\overline z_3^{a_2}\ ,~~~~{\rm
labels}~
~[a_1]_{\frac{2a^{}_2+a_1+3}{6}}\cr
&+& lower ~~weights ,
\eea
where $[\dots]$ are the  $SU(2)$ Dynkin labels. The lower weight states in the same Euler triplet can be obtained by repeated action of the lowering operator 

$$L_1-iL_2= \theta^{}_2\frac{\partial}{\partial \theta^{}_1}+ (z^{}_2\part1-\overline z^{}_1\partb2)\ ,$$
giving us all the states within each  the Euler triplet. Each triplet differs from the next by the degree of homogeneity. This implies that it is possible to move across the triplets by simple multiplication or differentiation. This is the object of the next section. The general solution to Kostant's equation is obtained by summing over the labels $a_1$ and $a_2$ and multiplying every term with a space-time field.

\setcounter{equation}{0}
\section{Relating the Triplets}
 In our notation, all Euler triplets can be written in terms of superfields with components of specific dependence on  the internal $z_i$ and $\overline z_i$ variables: 

\begin{itemize}
\item The  $\{0,0\}$  Euler triplet is described by the  superfield

\be
\Psi_{\{0,0\}}^{}=a^{}_0+a^{}_1\theta^{}_1+a^{}_2\theta^{}_2+a^{}_{12}
\theta^{}_1\theta^{}_2\ ,\ee
where the $a$'s do not depend on the internal $z$ variables, only on 
the center of mass variables $y^-$, and $x_i$, cf. (\ref{Ch}). Kinematic supersymmetry 
acts on its components by means of the operators

\be
{\cal Q}_+^m\approx -2p^+\theta_m\ ,\qquad \overline{\cal
Q}_+^m\approx 
\frac{\partial}{\partial\theta_m}\ ,\ee
so that this superfield describes  the $N=2$ Hypermultiplet on  the light-cone.

\item The   superfield

$$
\Psi^{}_{\{1,0\}}=b^{}_0z_3^{}+b^{}_1z^{}_1\theta^{}_1+b^{}_2(z^{}_1
\theta^{}_2+z^{}_2\theta^{}_1)+b^{}_3z^{}_2\theta^{}_2+\theta^{}_1
\theta^{}_2(b^{}_{12}z^{}_1+b^{'}_{12}z^{}_2)\ ,$$ 
where the $b$'s depend only on the center of mass variables, describes the $\{1,0\}$ triplet.

\item The $\{2,0\}$ Euler triplet is represented by

\bea\nonumber
\Psi_{\{2,0\}}&=&c^{}_0z_3^2+c_1^{}z_1^2\theta_1^{}+c_2^{}(z_1^2\theta_2^{}+2z_1^{}z_2^{}\theta^{}_1)+
c^{}_3(\theta_1^{}z_2^2+2z^{}_1z^{}_2\theta_2^{})+\\~&~&+~c^{}_4z^2_2\theta^{}_2+
(c^{}_{12}z_1^2+c^{\prime}_{12}z_1^{}z_2^{}+c^{\prime\prime}_{12}z_2^2)\theta_1^{}\theta_2^{}
\nonumber\ , \eea
with the $c$'s depending  on the center of mass variables as above. 
\end{itemize}
While there is no supersymmetry among the components of the higher Euler triplets, 
there are some supersymmetry-like operations which relate members of the Euler triplets, which we call {\it shadow supersymmetry}.

To further study algebraic operations that relate these superfields, it is convenient to introduce  the projection operators  onto the components of the chiral superfields:
\bea\nonumber
P^{}_0&=&(1-{\cal P}^{}_1)(1-{\cal P}^{}_2)    \ , 
\qquad
P^{}_1~=~{\cal P}^{}_1(1-{\cal P}^{}_2)\
,\\ \nonumber
P^{}_2&=&{\cal P}^{}_2(1-{\cal P}^{}_1)\
,\qquad
P^{}_{12}~=~{\cal P}^{}_1{\cal P}^{}_2 \ ,\\ \nonumber
\eea
where ${\cal P}_i\equiv \theta_i\partial/\partial\theta_i$, $i=1,2$. 
They satisfy the relations

\be P_aP_b=\delta_{ab}P_a\ .\ee

Our first task is to construct spectrum generating ladder  operators which, when applied to $\Psi_{\{a,0\}}$,
generate all components of $\Psi_{\{a+1,0\}}$. Since the non-Grassmann component for each triplet differs simply by the power in $z_3$, we start with  the  operator

\be A^{\dagger}_0~\equiv~z^{}_3~P^{}_0\ ,\ee
which multiplies the lowest component of any superfield by
$z_3$. It is  an $SU(2)$ singlet but carries  $-1/3$ helicity, and  acts
as a raising operator between Euler triplets. 
The corresponding lowering operator is

\be A^{ }_0~\equiv~~P^{}_0\frac{\partial}{\partial z^{}_3}\ .\ee
These two operators act as harmonic oscillators on the lowest
superfield component

\be
[A^{}_0~,~A^\dagger_0]~=~P^{}_0\ .\ee

The $\theta$-dependent terms of  two adjacent Euler triplets 
differ  by single powers of $z_1$ or $z_2$. Hence we are led to consider the 
$SU(2)$-doublet operator  
$z_i~(1-P^{}_0)$, where $(1-P_0)$ is the projection operator onto the $\theta$-dependent terms. 
When applied to the lowest Euler triplet, it generates some of the
states in $\Psi_{\{1,0\}}$, and more, as it fails to single out the  requisite
combinations of $z$'s and $\theta$'s. For this we need  the  operator

\be
{\cal P}~=~\frac{1}{2}\Big(1+\sum_{i,j=1,2}~z^{}_i\theta^{}_j
\frac{\partial}{\partial z^{}_j}\frac{\partial}{\partial\theta^{}_i}\Big)\ .\ee
It carries no helicity and is a singlet of $SU(2)$, as we  can  see by rewriting it in the form
\be
{\cal P}~=~\frac{1}{2}\Big(1+2\vec S\cdot \vec T + \frac{1}{2}\theta_i\frac{\partial}{\partial\theta_i}(z_j\frac{\partial}{\partial z_j}+
\overline z_j\frac{\partial}{\partial \overline z_j})
\Big)\ .\ee
Acting on the terms linear in $\theta$'s, it reduces simply to  

\be
{\cal P}~=~\frac{1}{2}+\vec S\cdot \vec T+  J/2\ ,\ee
while it is equal to one-half on the $1\ ,\, \theta_1\theta_2$ terms. With the help of 

\be
(\vec S\cdot\vec T)^2~=~-\frac{1}{2}\,\vec S\cdot\vec T+\frac{1}{3}\,\vec S\cdot\vec S\,J(J+1)\ ,\ee
we see that on the terms linear in $\theta$'s, 

\be
{\cal P}^2_{}~=~(J+\frac{1}{2})\,{\cal P}\ .\ee
so that
\be
{\cal M}~=~\frac{{\cal P}}{J+\frac{1}{2}}\ ,\ee
is a true projection operator on the $\theta$-linear terms. It is easy to check that 

\bea\nonumber
{\cal P}~z_1\theta^{}_1~=~z^{}_1\theta^{}_1\ ~~&;&~~ {\cal
P}~z^{}_2\theta^{}_2
~=~z^{}_2\theta^{}_2\ ;\\ {\cal
P}~(z^{}_2\theta^{}_1+z^{}_1\theta^{}_2)
&=&(z^{}_2\theta^{}_1+z^{}_1\theta^{}_2)\ ,\eea
while  the unwanted combination is annihilated

\be
{\cal P}~(z_2\theta_1-z_1\theta_2)~=~0\ .\ee 
Multiplying the $\theta$-dependent superfield components in the
$SU(2)$ spin $j$ representation by  $z_i$,
produces two $SU(2)$ representations,  $j+1/2$ and 
$j-1/2$. The action of ${\cal P}$ is to project out the latter (like the well-known projection that appears in $L-S$ coupling problems). Hence
the $SU(2)$-doublet ``creation" operators   have the simple form

\be A^{\dagger}_i~\equiv~ {\cal P}(1-P^{}_0) ~z^{}_i \ .\ee
Explicit computations yield 

\bea
A^{\dagger}_0~\Psi_{\{0,0\}}~&=&~a_0z^{}_3\nonumber\\
A^{\dagger}_1~\Psi_{\{0,0\}}~&=&~a^{}_1z^{}_1\theta^{}_1+\frac{a^{}_2}{2}
(z^{}_1\theta^{}_2+z^{}_2\theta^{}_1)
+a^{}_{12}z^{}_1\theta^{}_1\theta^{}_2
\ ,\nonumber\\
A^{\dagger}_2~\Psi_{\{0,0\}}~&=&~\frac{a^{}_1}{2}(z^{}_1\theta^{}_2+z^{}_2
\theta^{}_1)+a^{}_2z^{}_2\theta^{}_2
+a^{}_{12}z^{}_2\theta^{}_1\theta^{}_2
\ ,\nonumber\eea
so that their action on  the lowest Euler triplet generates all the
states in $\{1,0\}$, although with some redundancy. 
A similar construction holds for the  step-down operators, viz

\be 
A^{}_i~\equiv~ {\cal P}(1-~P^{}_0)\frac{\partial}{\partial z^{}_i} \ .\ee
Obviously, they annihilate the lowest Euler triplet

\be
A^{}_0~\Psi_{\{0,0\}}~=~A^{}_1~\Psi_{\{0,0\}}~=~A^{}_2~\Psi_{\{0,0\}}~=~0\ .\ee

The double application of the step-up operators  on the lowest Euler triplet,
\bea
(A^\dagger_0)^2~\Psi_{\{0,0\}}&=&a^{}_0z^2_3\ ,\nonumber\\
(A^\dagger_1)^2~\Psi_{\{0,0\}}&=&\frac{3}{2}a^{}_1z^{2}_1\theta^{}_1+a_2^{}(z^2_1\theta^{}_2
+2z^{}_1z^{}_2\theta^{}_1)+\frac{3}{2}a^{}_{12}z^2_1\theta^{}_1\theta^{}_2\ ,\nonumber\\~
(A^\dagger_2)^2~\Psi_{\{0,0\}}&=&a_1^{}(z^2_2\theta_1^{}+2z^{}_1z^{}_2\theta^{}_2)
+\frac{3}{2}a_2^{}z^2_2\theta^{}_2+\frac{3}{2}a^{}_{12}z^2_2\theta^{}_1\theta^{}_2\ ,\nonumber\\
A^\dagger_1~A^\dagger_2~\Psi_{\{0,0\}}&=&
\frac{1}{2}a_2^{}(z^2_2\theta_1^{}+2z^{}_1z^{}_2\theta^{}_2)+a^{}_1(z_1^2
\theta^{}_2+2z^{}_1z^{}_2\theta^{}_1)+\frac{3}{2}a^{}_{12}z^{}_1z^{}_2\theta^{}_1\theta_2^{}\
,\nonumber\\
A^\dagger_2~A^\dagger_1~\Psi_{\{0,0\}}&=&\frac{1}{2}a^{}_1(z_1^2
\theta^{}_2+2z^{}_1z^{}_2\theta^{}_1)+a_2^{}(z^2_2\theta_1^{}+2
z^{}_1z^{}_2\theta^{}_2)+\frac{3}{2}a^{}_{12}z^{}_1z^{}_2\theta^{}_1\theta_2^{}\ ,
\nonumber\eea
generates only the states in the $\{2,0\}$ multiplet, although the same states can be generated by different sets of operators. This construction  generalizes easily to all triplets of the form
$\{a_1,0\}$: acting on  any triplet $\Psi_{\{a_1,0\}}$, the step-up operators $A_{0,1,2}^\dagger$  
yield all the states in  the three $SU(2)$ representations of the triplet $\{a_1+1,0\}$.

In a completely analogous way we can find ``ladder''operators that can take us between the various levels $\Psi_{\{0,a_2\}}$. The first level is 

$$
\Psi^{}_{\{0,1\}}={\bar b}^{}_0{\bar z}_2 - {{\bar b}^{'}}_0{\bar z}_1+{\bar b}^{}_1{\bar z}^{}_2\theta^{}_1+{\bar b}^{}_2({\bar z}^{}_2
\theta^{}_2-{\bar z}^{}_1\theta^{}_1)-{\bar b}^{}_3{\bar z}^{}_1\theta^{}_2+{\bar b}^{}_{12}{\bar z}{}_3\theta^{}_1\theta^{}_2 ,$$ 
where the $\bar b$'s depend only on the center of mass variables as above.

The ladder operators can now be defined as

\be {\bar A}^{\dagger}_i~\equiv~ {\bar {\cal P}}~{\bar z}^{}_i \ ,\ee

where
\be {\bar {\cal P}}= (1- \epsilon_{ij}\epsilon_{kl}{\bar z}^{}_i\theta^{}_k\frac{\partial}{\partial {\bar z}^{}_l}
\frac{\partial}{\partial \theta^{}_j})
\ee  
and 
\be {\bar A}^{\dagger}_{12}~\equiv~P^{}_{12}{\bar z}^{}_3 \, \ee
and the ``annihilation''operators correspondingly as
\be 
{\bar A}_i~\equiv~ {\bar {\cal P}}(1-P^{}_{12})~{\frac{\partial}{\partial {\bar z}^{}_i}} \ ,\ee
and
\be 
{\bar A}_{12}~\equiv~P^{}_{12}~{\frac{\partial}{\partial{\bar z}^{}_3}} \, 
\ee

With these operators we have the neccessary means to step up and down in the full superfield. We can in this way find all the triplets. In the next section we study how to move within the triplets. Note that we have not specialized to the triplets with (half-)integer spins but kept the discussion general.

\section{Shadow Supersymmetry}
While ``naked'' supersymmetry acts on the components of the lowest
Euler multiplet, we can define ``shadow'' supersymmetry operations on
the components of $\Psi_{\{1,0\}}$ through the use of the step
operators. Define the operators

\be
{\cal Q}^{~[i]~}_{a~~b}~\equiv~ A^\dagger_a~\theta^{}_i~
A^{}_b\ ,\ee
where $a,b=0,1,2$, and $i=1,2$, and their inverses
\be
\overline{\cal Q}^{~[i]~}_{a~~b}~\equiv~ A^{\dagger}_a~\frac{\partial}{\partial\theta^{}_i~}
A^{}_b\ .\ee

The idea here is that we first step down and perform a supersymmetry on the lowest multiplet and then step up again. Since the step operators are well defined and keeps us in the superfield these operations will close on the  $\{1,0\}$ triplet. The anticommutator will typically yield terms of the form $z_i\frac{\partial}{\partial z^{}_i}P_j$. The anticommutator will also close on higher Euler triplets since again the step procedure is well defined.

Similarly we can define the operators

\be
{Q}^{~[i]~}_{a~~b}~\equiv~{\bar A}^\dagger_a~\theta^{}_i~
{\bar A}^{}_b\ ,\ee
where $a,b=0,1,2$, and $i=1,2$, and their inverses
\be
\overline{Q}^{~[i]~}_{a~~b}~\equiv~ {\bar A}^{\dagger}_a~\frac{\partial}{\partial\theta^{}_i~}
{\bar A}^{}_b\ .\ee

They operate in a similar fashion on $\Psi_{\{0,1\}}$ and indeed on the whole superfield. We can extend these shadow supersymmetries to operations that step up N steps 

\be
{\cal Q}^{~[i]~}_{a_1~a_2..a_N~~b_1~b_2...b_N}~\equiv~ A^\dagger_{a_1}~A^\dagger_{a_2}..A^\dagger_{a_N}~\theta^{}_i~
A^{}_{b_1}~A^{}_{b_2}...A^{}_{b_N}\ ,\ee
where $a_i,b_j=0,1,2$, and $i=1,2$, and their inverses
\be
\overline{\cal Q}^{~[i]~}_{a_1~a_2..a_N~~b_1~b_2...b_N}~\equiv~ A^{\dagger}_{a_1}~ A^{\dagger}_{a_2}... A^{\dagger}_{a_N}~\frac{\partial}{\partial\theta^{}_i~}
A^{}_{b_1}~A^{}_{b_2}..A^{}_{b_N}\ee
and similarly for the ones with $\bar A$. The anticommutators between these operators will again close to terms of the form $z_i\frac{\partial}{\partial { z}^{}_i}P_j$.

We can also check the anticommutators $$\{{\cal Q}^{~[i]~}_{a_1~a_2..a_N~~b_1~b_2...b_N}, 
\overline{\cal Q}^{~[i]~}_{a_1~a_2..a_M~~b_1~b_2...b_M}\}$$ to see that they close to step operators.

We have hence found a an infinite superalgebra for which the superfield is a representation. We have not been able to to write it, though, in a compact form. However, we want to stress that the representation has those marvellous properties that we described earlier.

We have here treated the general case. If we want to specialize to the triplets with (half)-integer helicities we have to demand powers of the step operators such that we only step between those triplets.

\setcounter{equation}{0}
\section{Discussion and conclusions}
In this report I have given an explicit solution to Kostant's equation leading to an infinite extension of the spectrum of the $N=2$ hypermultiplet. I have shown how one can define operators that take us between any states in the infinite-dimensional superfield defining a extended superalgebra. This solution defines a kinetic term for the supermultiplet. The next step for this action would be to try to find an interaction which carries the same symmetry as the kinetic term. In principle this could be tried in the same way as interaction terms are constructed in the light cone frame as nonlinear terms in the dynamic generators. However, we should remember that no interaction term has been found for the hypermultiplet. A further direction of the study would be to try to go back to a covariant formalism. There is no standard way though to do it in the light cone frame so it is difficult to say how hard such a problem is. We should remember though that this is a just a simpler example than the case we really want to study, $F_4/SO(9)$ and it is not clear how valuable these further studies are.
\vspace{5mm}

\noindent {\bf Acknowledgments}:
I wish to congratulate  Marc Henneaux to the very well deserved Francqui Prize and to thank him and Alexander Sevrin for their kind invitation to this exceptional meeting. I would also like to thank Prof.  L. Eyckmans and the Francqui Foundation for the hospitality extended to me at several occasions. I find the Francqui Foundation be a role model for how a foundations should work to promote excellent science. This work is  done in collaboration with Pierre Ramond. 


\vspace{5mm}



\begin{thebibliography}{99}
\bibitem{P1} T. Pengpan and P. Ramond,{\em Phys. Rep.} {\bf 315} 137(1999   
\bibitem{Curtwrong}T. Curtright, {\it Phys. Rev. Lett.} {\bf 48}, 1704(1982)
\bibitem{GKRS} B. Gross, B. Kostant, P. Ramond, and S. Sternberg, {\it

Proc. Natl. Acad. Scien.}, 8441 (1998)
\bibitem{BR} Lars Brink, P. Ramond, {\it Dirac Equations, Light-Cone Supersymmetry, and  Superconformal Algebras}
    In  Shifman, M.A. (ed.): The many faces of the superworld 398-416;HEP-TH 9908208. 
\bibitem{BR2}Lars Brink, {\it Euler Multiplets, Light-Cone Supersymmetry and
Superconformal Algebras}, in Proceedings of the International Conference
on Quantization, Gauge Theory, and Strings: Conference Dedicated to the
Memory of Professor Efim Fradkin, Moscow 2000.
\bibitem{PR2} P. Ramond, {\it Boson Fermion  Confusion: The String Path to Supersymmetry}, 
{\it Nucl.Phys.Proc.Suppl.} {\bf 101}, 45(2001); Hep-Th 0102012.
\bibitem{KOS} B. Kostant, ``{\em A Cubic Dirac Operator and the Emergence of Euler Number Multiplets of Representations for Equal Rank Subgroups}", {\it  Duke J. of Mathematics} {\bf 100}, 447(1999). 
\end{thebibliography}
\end{document}